\documentstyle[12pt,graphics]{article}

\title{Can the flyby anomalies be explained by a modification of inertia?}

\author{M.E.~McCulloch \footnote{Affiliation pending, Devon, UK, memcculloch@btinternet.com}}

\begin{document}

\maketitle

\section*{ABSTRACT}

The flyby anomalies are unexplained velocity increases of 3.9, 13.5, 0.1 and 
1.8~mm/s observed near closest approach during the Earth flybys of the 
Galileo, NEAR, Cassini and Rosetta spacecraft. Here, these flybys are 
modelled using a theory that assumes that inertia is caused by a form of
Unruh radiation, modified by a Hubble-scale Casimir effect. This theory 
predicts that when the craft's accelerations relative to the galactic centre 
approached zero near closest approach, their inertial masses reduced 
for about 10$^{-7}$s causing Earthward jumps of 2.6, 1.2, 1.4 and 1.9 mm/s 
respectively, and, to conserve angular momentum, increases in orbital 
velocity of a few mm/s that, except NEAR's, were quite close to those 
observed. However, these results were extremely sensitive to the Hubble 
constant used. As an experimental test of these ideas, it is proposed that 
metamaterials could be used to bend Unruh radiation around objects, 
possibly reducing their inertial mass.

\newpage

\section{Introduction}

Several large-scale dynamical anomalies remain unexplained, including: 1) the 
galaxy rotation problem noticed by Zwicky~(1933) in which the stars in 
galaxies are too energetic to be bound by standard theories of gravity, 
2) the Pioneer anomaly discovered by Anderson $et~al.$~(1998) in which the 
Pioneer craft seem to be attracted to the Sun slightly more than expected 
and 3) the flyby anomalies described by Antreasian and Guinn~(1998) and Anderson 
$et~al.$~(2006) in which some spacecraft during gravity assist flybys have shown 
anomalous velocity jumps of a few mm/s. These problems can all be interpreted 
as unexpected increases in gravitational interaction (either an increase in 
Newton's gravitational constant $G$, an increase in gravitational mass, or 
a loss of inertial mass), and the first two anomalies appear at very low 
accelerations (of the order of 10$^{-10}ms^{-2}$).

An increase of $G$ for such low accelerations is the approach of some versions of MOND 
(MOdified Newtonian Dynamics), an empirical theory introduced by Milgrom~(1983) 
to solve the galaxy rotation problem. The problem with this approach is that it 
does not explain why the Pioneer craft and the planets behave differently. An 
increase of gravitational mass is the approach of the popular dark matter 
hypothesis of Zwicky~(1933). However, it is possible to fit dark matter 
distributions to solve the problem, but no theory yet exists to explain 
the distributions. It is also difficult to explain the Pioneer anomaly 
using dark matter since, again, the planets would also be effected.

The third approach: reducing the inertial mass for low accelerations, was 
suggested by Milgrom~(1999) who realised that MOND could be interpreted this 
way. As he noted, there are observations that imply that it is inertia that 
should be modified. For example: the possible change in behaviour of the 
Pioneer craft upon moving from a bound to an unbound trajectory (to be confirmed, 
or not, soon, by the Pioneer team, see Toth and Turyshev, 2006), and the planets, 
which are on bound orbits, 
do not seem to show the anomaly. A modification of inertia is the only approach 
that can be made to depend upon trajectory, as this phenomenon appears to. The 
problem with this approach is that it violates the equivalence principle. However, 
as noted by McGaugh~(2007) this principle has not been tested at very low 
accelerations, which are difficult to attain on Earth.

A possible model for modified inertia can be found by starting from the work of
Haisch $et~al.$~(1994) who suggested that the inertial mass of a body is caused 
by a drag from a form of Unruh radiation (Unruh,~1954) which is only apparent 
upon acceleration. Since the wavelength of this radiation lengthens as the 
acceleration reduces Milgrom~(1994,~1999) suggested that there would be a break 
in this quantum vacuum effect and a consequent loss of inertia for very low 
accelerations when the Unruh wavelengths exceed the Hubble distance showing
behaviour similar to MOND, though the behaviour for intermediate accelerations, 
like that of the Pioneer craft, was undefined.

McCulloch~(2007) proposed a model which could be called Modified 
Inertia due to a Hubble-scale Casimir effect (MiHsC), in which, as the acceleration 
reduces the radiation is diminished linearly since fewer wavelengths fit 
within twice the Hubble distance, a more gradual process than Milgrom's break. 
In this model, the equivalence principle ($m_i$=m$_g$) becomes
\begin{equation}
m_I = m_g \left( 1- \frac{\beta \pi^2 c^2}{a \Theta} \right).  \label{eq:mi}
\end{equation}
where $m_I$ is the modified inertia, $m_g$ is the mass of the spacecraft, 
$\beta=0.2$ (from the empirically-derived Wien's constant), $c$ is the speed 
of light, $\Theta$ is twice the Hubble distance $2c/H$, and $a$ is the acceleration 
of the craft relative to the galactic centre. This model agreed with the Pioneer 
anomaly beyond 10~au from the Sun without the need for adjustable parameters 
(McCulloch,~2007). However, the model also predicted an anomaly within 10~au, 
where none was observed (MiHsC may not be applicable to bound trajectories).

During four Earth gravity assist flybys, anomalous velocity increases of a few mm/s 
were observed near closest approach (Antreasian and Guinn,~1998, 
Anderson~$et~al.$,~2007) and these are known as the flyby anomalies. 
Table 1 lists the spacecraft involved (column 1), the flyby dates (column 2) 
and the anomalous velocity increase seen (dV, column 3). So far, no explanations 
for these events have been found (Anderson~$et~al.$,~2007). 

It may seem that the flyby anomalies are unlike the galaxy rotation problem 
and the Pioneer anomaly, because the accelerations involved are larger. 
However, Ignatiev~(2007) suggested that small short-lived zones of modified 
inertia could occur on Earth, under very rare conditions in which the $total$ 
acceleration relative to the galactic centre approached zero. It is reasonable 
to ask whether this occurs for near-Earth objects too. In this paper the 
flybys are modelled using the version of modified inertia suggested by 
McCulloch~(2007), in an attempt to explain the flyby anomalies.

\begin{center}
\begin{table}[htp]
\begin{center}
\begin{tabular}{l|c|c}                       \hline
Mission & Flyby date & Observed dV           \\
        &            & (mm/s)                \\ \hline\hline
Galileo & 8/12/90 & \bf{3.92}  $\pm 0.08$     \\ \hline
NEAR    & 23/1/98 & \bf{13.46} $\pm 0.13$     \\ \hline
Cassini & 18/8/99 & \bf{0.11}                 \\ \hline
Rosetta & 4/3/05  & \bf{1.82}  $\pm 0.05$     \\ \hline
\end{tabular}
\end{center}
\caption{Observed flybys with good data (Anderson~$et~al.$~2007, 
L\"ammerzahl~$et~al.$, 2006). The error
bars are also shown, where known.}
\end{table}
\end{center}

\section{Method}

Ephemeris data for the Earth, Moon and the spacecraft, for each of the four 
flybys were downloaded at 1 minute temporal resolution from the excellent
JPL Horizons website for the dates shown in table 1, column 2. The spacecraft 
trajectories were then modelled using Newtonian dynamics and MiHsC.

A starting point along the trajectory about 1 hour before closest approach was
chosen and the model was initialised with the position and velocity data from
JPL at that point and then run for six hours with a time step of 0.2 s. This 
time step was increased to O(10$^{-7}$s) during the time-step within which the
acceleration passed through zero to better resolve the decrease in inertial 
mass predicted by equation (1). During this sub-model phase the acceleration 
was forced to remain above $6.9 \times 10^{-10}ms^{-2}$ since in MiHsC the 
acceleration cannot pass below this minimum value. This is because for very 
low accelerations the inertial mass reduces, and this increases the acceleration 
again. An equilibrium is established at this value (see McCulloch, 2007). 
The numerics were handled using a simple forward-stepping scheme. The value
used for the Hubble constant was $2.33 \times 10^{-18}s^{-1}$. In McCulloch (2007)
the value used was $2.3 \times 10^{-18}s^{-1}$. This difference is discussed below.

The inertial masses of the Earth, Moon, Sun and spacecraft were not alterred in 
this simulation because it was assumed that bound orbits do not show an anomalous 
effect, using the example of the Pioneer craft which did not appear to show an 
anomaly while bound to the Sun (to be confirmed, or not, soon: see Toth and 
Turyshev, 2006) and the planets, which, of course, do not show an anomaly.

Figures 1-4a (the top left plots) show the trajectories for each of the flybys 
looking down on the Earth's north pole. In each plot the Earth's trajectory is in 
bold, the moon's is lighter and the spacecraft's is shown lighter still and 
shows a change in direction near close approach due to the Earth's gravity.

\section{Results}

Figure~1 shows the results for the Galileo flyby which occured on the 8th of
December, 1990. As discussed above the top left plots show the trajectories.
The bottom left plot shows the heliocentric spacecraft velocity (see the upper
curve and the left hand axis). The peak velocity occured about one hour into 
the run at closest approach, and was about 37~km/s in this case. Note that the
final velocity is greater than the initial, because this flyby added momentum
to the craft. The bottom curve shows the total acceleration of the craft relative 
to the Sun (the axis is on the right). Ideally, the acceleration of interest is
that relative to the galactic centre (GC), but the Sun's acceleration relative 
to the GC is very small: about 10$^{-10}ms^{-2}$, so an origin at the Sun should
suffice. Near closest approach the acceleration passed close to zero, an event 
which reduces the inertial mass according to equation (1). The inertial mass is 
shown by the thicker horizontal line and its axis is on the right. The predicted 
reduction of inertial mass can be seen as a vertical spike. It reduced from 
2497~kg to 2.1~kg, but over a duration of only $3 \times 10^{-7}$s. The 
negative spike in inertial mass seen here is similar to the SHLEM (Static
High Latitude Equinox Modified inertia) effect predicted to occur on Earth 
on very rare occasions by Ignatiev~(2007).

The predicted reduction in inertial mass caused an increase in the acceleration 
towards the Earth and an Earthward velocity jump of 2.6~mm/s (see Table 2, 
column 3). To conserve angular momentum the orbital velocity then increased. 
The predicted anomalous increase in heliocentric velocity is shown by the solid 
line in the top right plot and the anomalous geocentric velocity is shown by the 
dotted line. As the craft jumped towards the Earth exchanging momentum with it, 
its geocentric velocity decreased briefly, but the extra orbital velocities 
increased quickly to about 2-3~mm/s and thereafter increased ever more slowly. 
They were still rising slightly after 6 hours, but the increase of interest is
the one that occurs, like the flyby anomalies, near closest approach. The bottom 
right plot shows the difference in the craft's distance from Earth between the 
MiHsC and Newtonian runs. In the MiHsC run, the Earthward jump in velocity at 
first decreased this distance, but the greater orbital velocity eventually leads 
to a distance anomaly which increases at a rate of 10~m per hour, or about 3 mm/s.

\begin{table}[htp]
\begin{center}
\begin{tabular}{l|c|c|c|c|c}                                                \hline
Mission & Observed  & Predicted    & Craft mass  & Mass loss          & Time after   \\
        & dV        & Earthward dV & (minimum)   & duration           & C.A.    \\
        & (mm/s)    & (mm/s)       &   (kg)      &   (s)           & (mins) \\ \hline\hline
Galileo & \bf{3.92}  & 2.6         &  2497 (2.1) & 3 $\times 10^{-7}$ & 12    \\ \hline
NEAR    & \bf{13.46} & 1.2         &   730 (0.6) & 0.6 $\times 10^{-7}$ & -2    \\ \hline
Cassini & \bf{0.11}  & 1.4         &  4612 (4.0) & 2 $\times 10^{-7}$ &  7    \\ \hline
Rosetta & \bf{1.82}  & 1.9         &  2895 (2.5) & 2 $\times 10^{-7}$ &  9    \\ \hline
\end{tabular}
\end{center}
\caption{Column 1: The flyby name, Column 2: The observed orbital velocity jumps. 
Column 3: the predicted $Earthward$ velocity jumps (mm/s). Column 4: the nominal 
and minimum inertial masses (kg). Column 5: the duration of the mass-loss event 
(in seconds). Column 6: the times of occurence of the mass loss event relative 
to closest approach (in minutes).}
\end{table}

The predicted orbital jumps in velocity for all four of the flybys are shown in 
Figures 1-4 (the top right plots) and summarised in table 2. In the table columns 
2 and 3 show the observed orbital and predicted $Earthward$ velocity jumps. The
predicted orbital velocity jumps are shown in Figures 1-4. The predicted velocity 
jumps for Galileo and Rosetta agreed quite well with those observed , but that 
for NEAR was an order of magnitude smaller. Since equation (1) is a difference 
between two terms, one of which contains the Hubble constant $H$, these results 
depended strongly on the value chosen for $H$ which, for these results, was 
$2.33 \times 10^{-18}s^{-1}$. For a value of $H=2.3 \times 10^{-18}s^{-1}$ the 
predicted jumps were less than 1~mm/s and for $H \geq 2.34 \times 10^{-18}s^{-1}$, 
the jumps were negative since the second term on the right hand side of equation 
(1) became dominant. Therefore, this theory succeeds for only a very narrow range of 
values of $H$. Column~4 of table~2 shows the gravitational mass (and normal inertial 
mass) of the craft and, in brackets, the minimum inertial mass achieved during 
the mass loss event. Column~5 shows the duration of that event. The loss of 
inertial mass was greatest for the NEAR flyby (it reduced from 730~kg
to 0.6~kg), but the duration of the mass loss was shorter. Column~6 of table~2 
shows the time, relative to close approach (CA), that the jump (and the mass 
loss event) occured. They occured a few minutes after closest approach (positive
values) except in the case of NEAR where the mass loss occured two minutes before. 
Unfortunately, the exact timing of the event can not be compared with these results
since contact with the craft was lost by the tracking stations near closest approach.

\section{An experimental test}

If the ideas discussed here are correct then it should be possible to reduce the 
inertial mass of an object by reducing the Unruh radiation is sees upon acceleration. 
One way to achieve this could be to use the metamaterials recently devised by 
Pendry~$et~al.$~(2006), or Leonhardt~(2006). They have demonstrated that 
radiation can be bent around an object that is smaller than that wavelength
using a metamaterial, in such a manner that the rays exit the vicinity of the
object in the same direction that they entered, so that the object becomes invisible 
at that wavelength. This also implies a cancelation of the momentum that would have 
been given to the object by the radiation. For an object with a typical acceleration 
of 9.8 ms$^{-2}$ the Unruh wavelength is about 10$^{16}$~m, which seems rather
large, but Pendry~$et~al.$~(2007) have proposed metamaterials that can bend the 
magnetic component of radiation with wavelengths even of this order.

Another way to think about this is that the bending of radiation around the
object can be arranged to create a boundary similar to the one considered
here to exist at the edge of the observable universe - the size of which is
the $\Theta$ in equation (1). In the examples here $\Theta$ was $2.6 \times 
10^{26}~m$, but for an object surrounded by a carefully arranged metamaterial 
$\Theta$ could be reduced in size, making the inertial drop predicted by 
equation (1) far more detectable.

\section*{Conclusions}

A model of modified inertia, using a Hubble-scale Casimir effect predicts 
orbital velocity jumps near closest approach which are of an similar order
of magnitude to the flyby anomalies, except for the NEAR flyby. 

The results are extremely sensitive to the Hubble constant, and the numerical 
schemes used to predict the trajectories were relatively simple. It is therefore 
hoped that other specialist groups can reproduce these results. 

As an experimental test for these ideas it is proposed here that newly-developed 
metamaterials may be used to reduce the impact of Unruh radiation on an 
accelerated object, thereby measurably reducing its inertial mass.

\section*{Acknowledgements}

This is a summary of a presentation given at a symposium at the British 
Interplanetary Society on 15th November 2007 and thanks for K.Long for 
organising this event. Thanks also to J.Anderson, J.Pendry and C.Smith for advice.

\section*{References}

Anderson,~J.D.,~J.K.~Campbell, M.M.~Nieto,~2007. The energy
transfer process in planetary flybys. $New~Astron.$, 12, 383-397.
astro-ph/0608087

Anderson,~J.D., P.A.~Laing, E.L.~Lau, A.S.~Liu, M.M.~Nieto and S.G.~Turyshev, 1998.
Indication, from Pioneer 10/11, Galileo and Ulysses Data, of an apparent weak
anomalous, long-range acceleration. $Phys.~Rev.~Lett.$, 81, 2858-2861.

Antreasian,~P.G. and J.R.~Guinn,~1998. Investigations into the unexpected
delta-v increases during the Earth gravity assists of Galileo and NEAR.
Paper no. 98-4287 presented at the AIAA/AAS Astrodynamics Specialist Conf.
and Exhibition, Boston, 1998.

Haisch,~B., Rueda,~A. and Puthoff,~H.E., 1994. Inertia as a zero-point field 
Lorentz force. $Phys.~Rev.~A$,~49,~678.

Ignatiev,~A.Y.,~2007. Is violation of Newton's second law possible.
$Phys.$ $Rev.$ $Lett.$, 98, 101101. gr-qc/0612159

L\"ammerzahl,~C, O.~Pruess and H.~Dittus,~2006. Is the physics within
the solar system really understood? gr-qc/0604052

Leonhardt,~U.,~2006. Optical conformal mapping. $Science$, 312, 1777-80.

McCulloch,~M.E.,~2007. Modelling the Pioneer anomaly as modified inertia.
$MNRAS$, 376, 338-342. astro-ph/0612599

McGaugh,~S.S.,~2007. Modified Newtonian Dynamics close to home (reply).
$Science$, Letters, 318, 568-570.

Milgrom,~M.,~1983. A modification of the Newtonian dynamics as a possible
alternative to the dark matter hypothesis. $ApJ$, 270, 365.

Milgrom,~M.,~1994. Ann.Phys., 229, 384.

Milgrom,~M.,~1999. The Modified Dynamics as a vacuum effect. 
$Phys.~Lett.~A$, 253, 273.

Pendry,~J.B, D.~Schurig and D.R.~Smith,~2006. Controlling electromagnetic
fields. $Science$, Vol.~312, 5781, 1780-1782.

Pendry,~J.B. and B.~Wood, 2007. Metamaterials at zero frequency.
$J.Phys:Condens.Matter$, 19, 076208.

Toth,~V.T. and S.G.~Turyshev, 2006. The Pioneer anomaly: seeking an
explanation in newly recovered data. gr-qc/0603016

Zwicky,~F., 1933. Helv. Phys. Acta, 6, 110

\newpage

\resizebox{400pt}{450pt}{\includegraphics{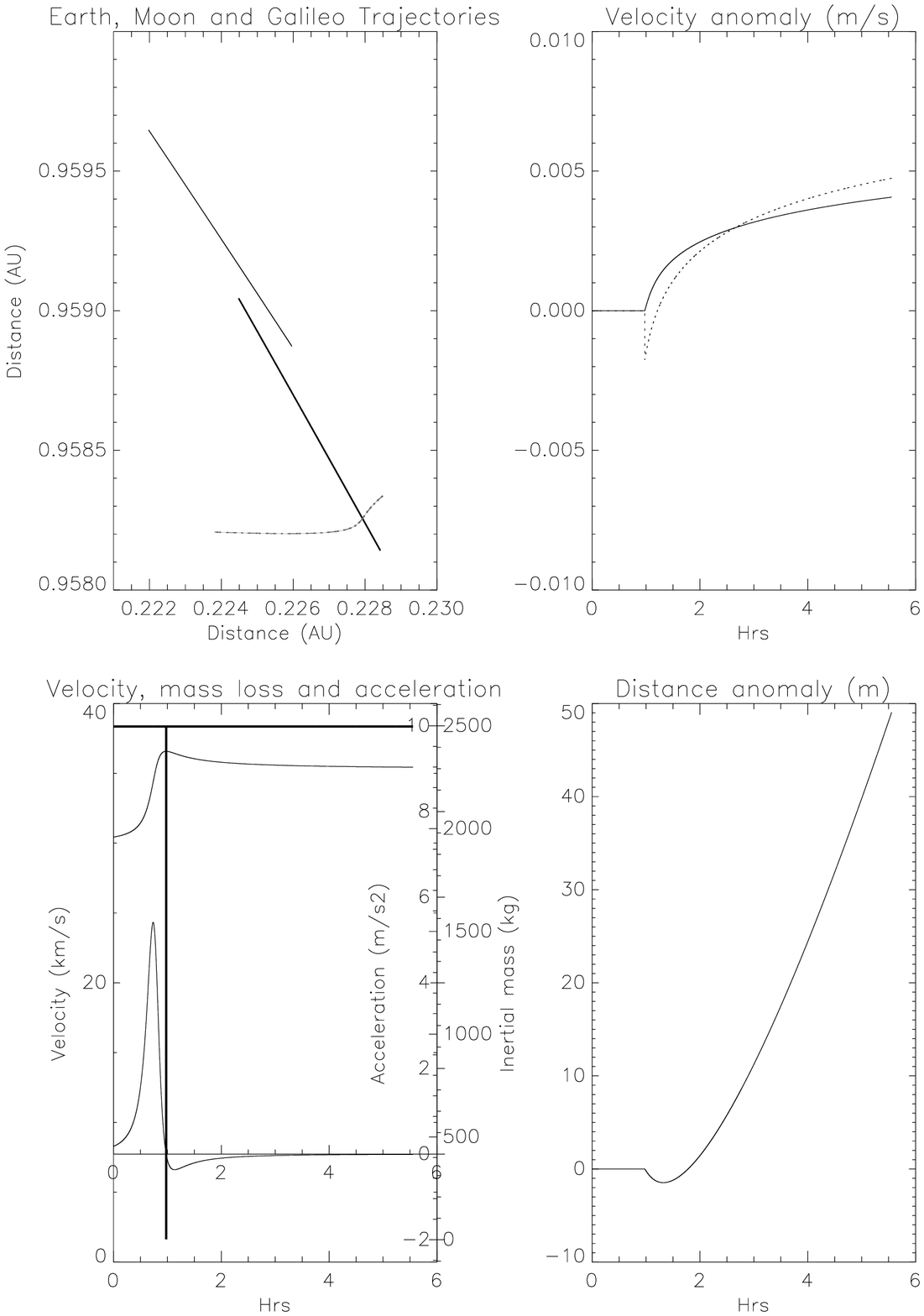}}

Figure 1. The Earth flyby by Galileo on the 8th December, 1990.
Top left: the Earth, Moon and spacecraft trajectories, bottom left: 
the velocity (km/s), acceleration (m/s2) and inertial mass (kg),
top right: the predicted velocity anomaly (m/s), bottom right: 
the predicted distance anomaly (m).

\newpage

\resizebox{400pt}{450pt}{\includegraphics{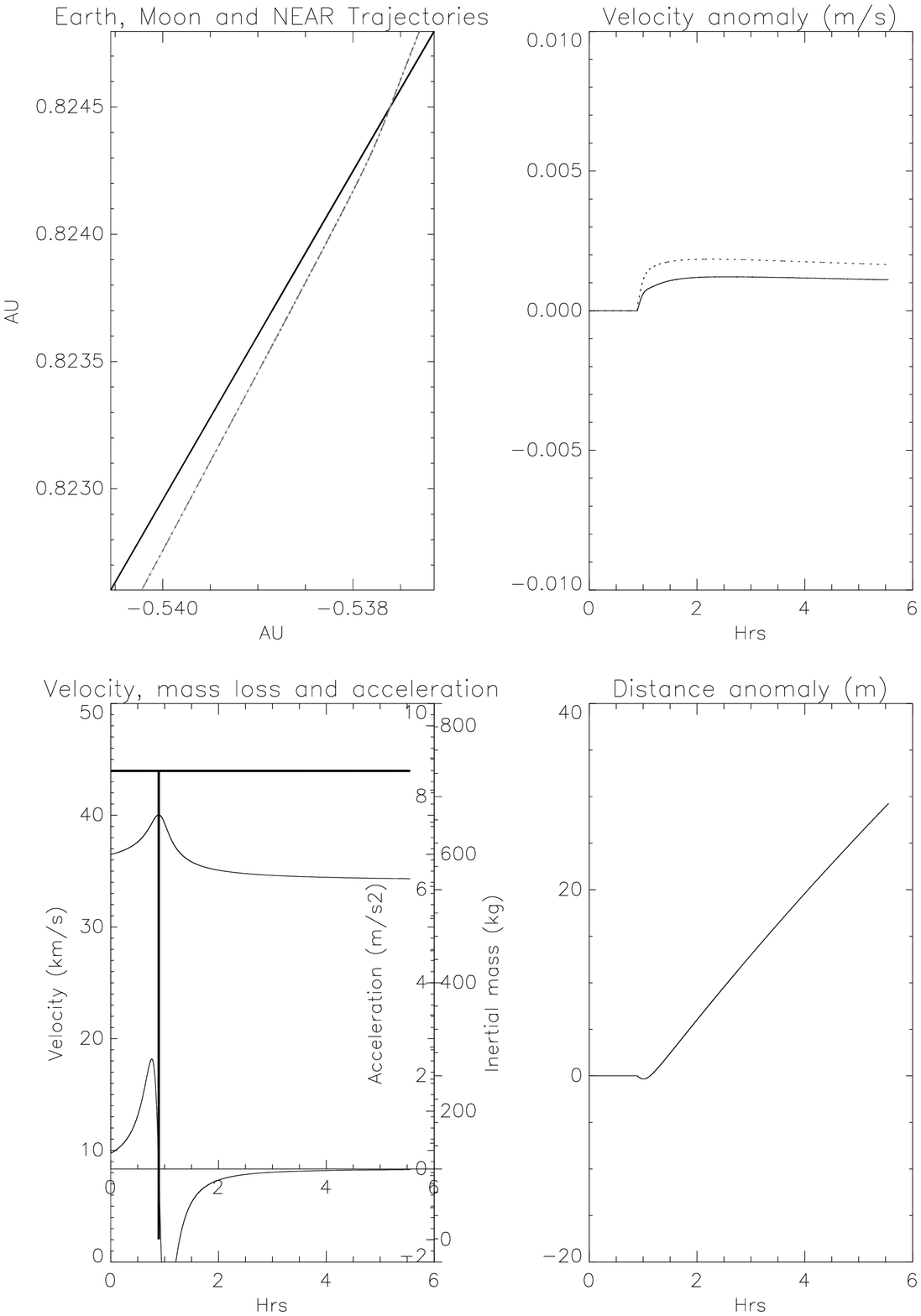}}

Figure 2. The Earth flyby by NEAR on the 23rd of January, 1998.
Top left: the Earth, Moon and spacecraft trajectories, bottom left: 
the velocity (km/s), acceleration (m/s2) and inertial mass (kg),
top right: the predicted velocity anomaly (m/s), bottom right: 
the predicted distance anomaly (m).

\newpage

\resizebox{400pt}{450pt}{\includegraphics{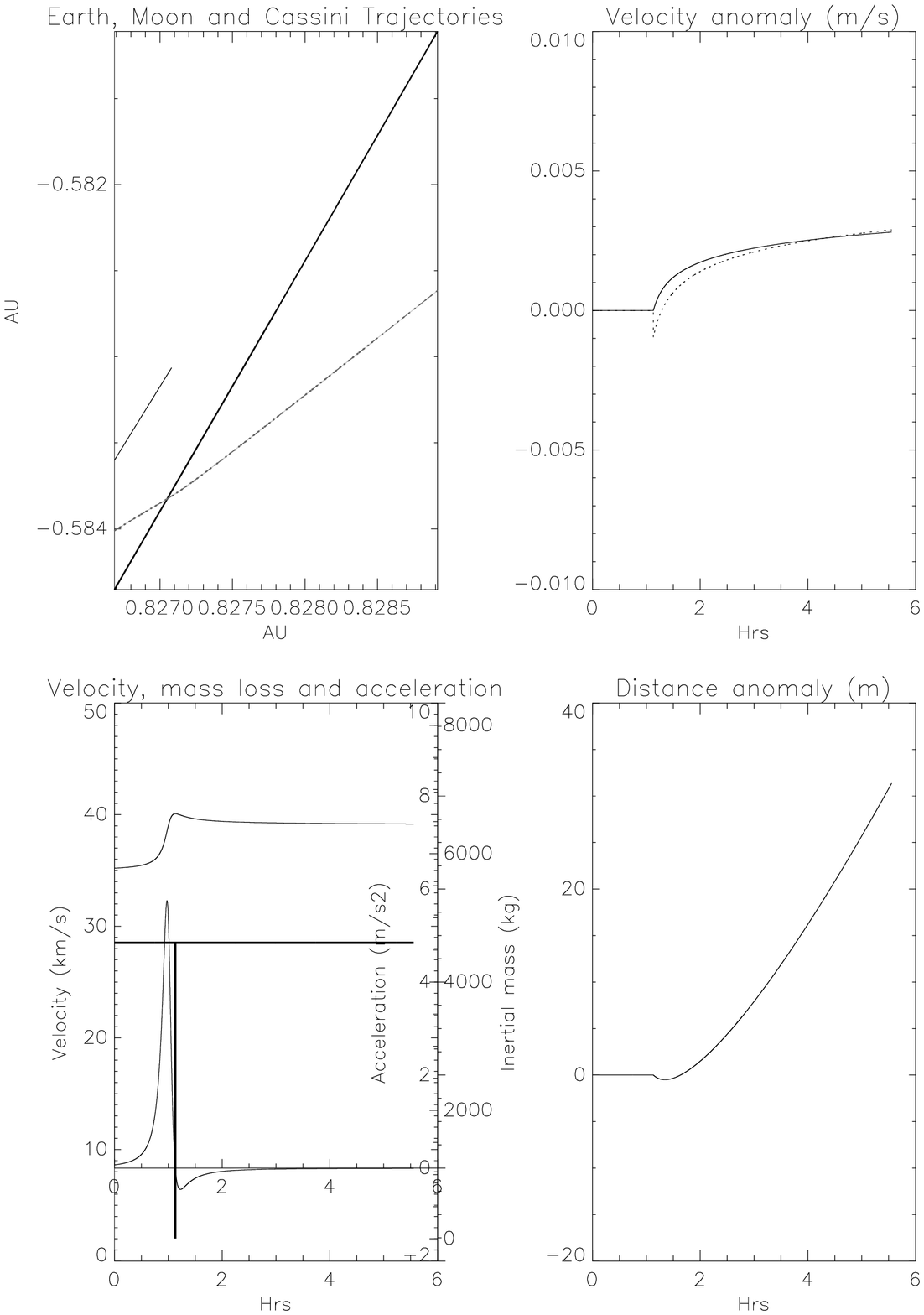}}

Figure 3. The Earth flyby by CASSINI on the 18th of August, 1999.
Top left: the Earth, Moon and spacecraft trajectories, bottom left: 
the velocity (km/s), acceleration (m/s2) and inertial mass (kg),
top right: the predicted velocity anomaly (m/s), bottom right: 
the predicted distance anomaly (m).

\newpage

\resizebox{400pt}{450pt}{\includegraphics{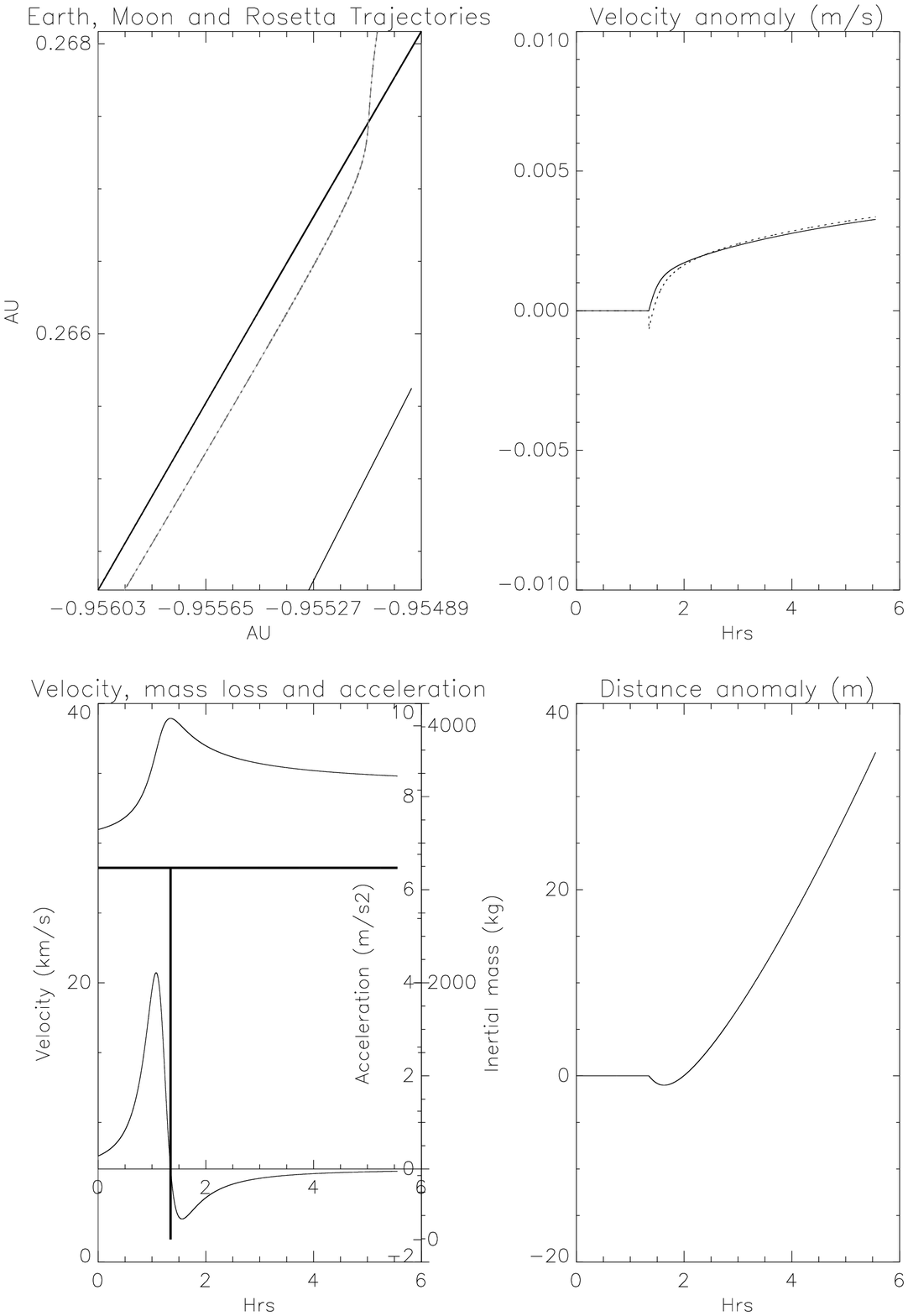}}

Figure 4. The Earth flyby by ROSETTA on the 4th of March, 2005.
Top left: the Earth, Moon and spacecraft trajectories, bottom left: 
the velocity (km/s), acceleration (m/s2) and inertial mass (kg),
top right: the predicted velocity anomaly (m/s), bottom right: 
the predicted distance anomaly (m).

\end{document}